\documentclass[conference,10pt]{IEEEtran}
\usepackage{cite}

\ifCLASSINFOpdf
  \usepackage[pdftex]{graphicx}
\else
\fi
\usepackage{color}
\usepackage{caption}
\usepackage{subcaption}

\usepackage[cmex10]{amsmath}
\usepackage{amsthm}
\usepackage{amssymb}

\usepackage{multirow}

\usepackage{stfloats}

\usepackage{threeparttable}

\newtheorem{theorem}{Theorem}
\newtheorem{proposition}{Proposition}

\newcommand{\figwidth}{0.44}

\begin{document}
%
\title{Faulty Successive Cancellation Decoding of Polar Codes for the Binary Erasure Channel}
%
%
%

\author{\authorblockN{Alexios Balatsoukas-Stimming and Andreas Burg}\\%
\authorblockA{Telecommunications Circuits Laboratory, EPFL, Lausanne, Switzerland.\\}%
E-mail: \{alexios.balatsoukas,andreas.burg\}@epfl.ch}%

%
%

\maketitle

\begin{abstract}
We study faulty successive cancellation decoding of polar codes for the binary erasure channel. To this end, we introduce a simple erasure-based fault model and we show that, under this model, polarization does not happen, meaning that fully reliable communication is not possible at any rate. Moreover, we provide numerical results for the frame erasure rate and bit erasure rate and we study an unequal error protection scheme that can significantly improve the performance of the faulty successive cancellation decoder with negligible overhead.
\end{abstract}


%
\IEEEpeerreviewmaketitle

\section{Introduction}\label{sec:introduction}
%
%
%
%
Parameter variation is expected to play a significant role in the design of integrated circuits in the nanoscale era \cite{Ghosh2010}. Therefore it will become more and more difficult to guarantee the correct behavior of integrated circuits at the gate level, meaning that the hardware may become \emph{faulty} in the sense that data is not always processed or stored correctly. Traditional methods to ensure accurate hardware behavior, such as using larger transistors or circuit-level error correcting codes, are costly in terms of both area and power. Fortunately, many applications are inherently fault tolerant in the sense that they do not fail catastrophically under faulty hardware. A good example of such an application are wireless communication systems, since the data is already probabilistic in nature due to transmission over a noisy channel. 

More specifically, faulty iterative decoding of LDPC codes was first studied in~\cite{Varshney2011}, where the Gallager A and sum-product algorithms are considered. Later studies also targeted the Gallager B algorithm~\cite{Yazdi2013,Leduc-Primeau2012} and the min-sum algorithm~\cite{Kameni2013,Balatsoukas2014}. All of the aforementioned studies provide valuable insight into the limitations of LDPC codes under various decoding algorithms and fault models. Unfortunately, in most cases, the conclusion is that fully reliable communication is not possible when faults are present inside the decoder itself.

Polar codes~\cite{Arikan2009}, constitute a different class of channel codes which has recently attracted significant attention, albeit not in the context of faulty decoding. Polar codes are provably capacity achieving over various channels and they have an efficient and structured successive cancellation decoding algorithm whose complexity scales like $O(N\log N)$, where $N$ is the length of the code. Moreover, encoding can also be performed with complexity that scales like $O(N\log N)$.

\subsubsection*{Contribution and Outline}
In this paper we provide an initial study of successive cancellation decoding of polar codes for transmission over the binary erasure channel (BEC) under a simple fault model. We show that, under the fault model assumed in this paper, fully reliable communication is no longer possible. Moreover, we provide numerical frame erasure rate (FER) results and we employ a fault-tolerance method which shows significant improvements with very low overhead. The remainder of this paper is organized as follows. Section~\ref{sec:background} provides some background on the construction and decoding of polar codes. In Section~\ref{sec:faulty}, we introduce the fault model that is used throughout this paper and we prove that fully reliable communication using polar codes is not possible under faulty decoding over the BEC. Section~\ref{sec:numerical} provides some numerical results on the frame erasure rate and the bit erasure rate and in Section~\ref{sec:protection} we study an unequal error protection scheme. Finally, Section~\ref{sec:conclusion} concludes this paper.

\section{Polar Codes}\label{sec:background}
Following the notation of \cite{Arikan2009}, we use $a_1^N$ to denote a row vector $(a_1,\hdots,a_N)$ and $a_i^j$ to denote the subvector $(a_i,\hdots,a_j)$. If $j < i$, then the subvector $a_i^j$ is empty. We use $\log(\cdot)$ to denote the binary logarithm. If $\mathcal{I}$ is a set of indices, then $a_{\mathcal{I}}$ denotes the subvector formed by taking the elements of $a_1^N$ whose indices belong to $\mathcal{I}$. We denote the binary erasure channel with erasure probability $p$ as BEC$(p)$.

\subsection{Construction of Polar Codes}
Let $W$ denote a binary input discrete and memoryless channel with input $u~\in~\{0,1\}$, output $y~\in~\mathcal{Y}$, and transition probabilities $W(y|u)$. A polar code is constructed by  applying a $2 \times 2$ \emph{channel combining} transformation recursively on $W$ for $n$ times, followed by a \emph{channel splitting} step~\cite{Arikan2009}. This results in a set of $N = 2^n$ channels, denoted by $W_N^{(i)}(y_1^N,u_1^{i-1}|u_i),~i=1,\hdots,N$. 

In principle, it is possible to compute the Bhattacharyya parameters $Z_i \triangleq Z\left(W_N^{(i)}(Y_1^N,U_ 1^{i-1}|U_i)\right),~i=1,\hdots,N$. In practice, finding an analytical expression turns out to be a very hard problem, except for the case of the BEC, where an exact recursive calculation is possible \cite{Arikan2009}. The construction of a polar code of rate $R \triangleq \frac{k}{N},~0 < k < N,$ is completed by choosing the $k$ channels with the lowest $Z_i$ as \emph{non-frozen} channels which carry information bits, while \emph{freezing} the remaining channels to some values $u_i$ that are known both to the transmitter and to the receiver. The set of frozen channel indices is denoted by $\mathcal{A}^c$ and the set of non-frozen channel indices is denoted by $\mathcal{A}$. The encoder generates a vector $u_1^N$ by setting $u_{\mathcal{A}^c}$ equal to the known frozen values, while choosing $u_{\mathcal{A}}$ freely. A codeword is obtained as $x_1^N = u_{1}^NG_N,$ where $G_N$ is the generator matrix.

\subsection{Successive Cancellation Decoding of Polar Codes}
The successive cancellation (SC) decoding algorithm~\cite{Arikan2009} starts by computing an estimate of $u_1$, denoted by $\hat{u}_1$, based only on $y_1^N$. Subsequently, $u_2$ is estimated using  $(y_1^N,\hat{u}_1),$ etc. Let the log-likelihood ratio (LLR) for $W_N^{(i)}(y_1^N,\hat{u}_1^{i-1}|u_i)$ be defined as 
\begin{align}
	L(y_1^N,\hat{u}_1^{i-1}|u_i) \triangleq \log \frac{W_N^{(i)}(y_1^N,\hat{u}_1^{i-1}|u_i=0)}{W_N^{(i)}(y_1^N,\hat{u}_1^{i-1}|u_i=1)}. 
\end{align}
Decisions are taken according to
\begin{align}
	\hat{u}_i & =\left\{ \begin{matrix} 0, & L(y_1^N,\hat{u}_1^{i-1}|u_i) > 0 \text{ and } i \in \mathcal{A}, \\ 1, & L(y_1^N,\hat{u}_1^{i-1}|u_i) < 0 \text{ and } i \in \mathcal{A} \\u_i, & i \in \mathcal{A}^c. \end{matrix} \right. \label{eq:scdec}
\end{align}
If $L(y_1^N,\hat{u}_1^{i-1}|u_i) = 0$, the decoder declares a failure. In order to calculate each $L(y_1^N,\hat{u}_1^{i-1}|u_i)$, the channel LLRs $\log \frac{W(y_i|x_i=0)}{W(y_i|x_i=0)}$ are combined through the stages of a decoding graph containing nodes of two types. Each node has two input LLRs and one output LLR, which we denote by $m_1, m_2,$ and $m$, respectively. For the first type of node, we have
\begin{align}
	m	& = m_1 + (-1)^{\hat{u}_s}m_2,
\end{align}
where $\hat{u}_s$ is called a \emph{partial sum} and it is always a \mbox{modulo-2} sum of some of the codeword bits that have already been decoded. For the second type of node, we have
\begin{align}
	m	& = 2 \tanh ^{-1}\left(\tanh(m_1/2)\tanh(m_2/2)\right).
\end{align}
Due to the similarity of these update rules with the update rules of the sum-product algorithm that is commonly used to decode LDPC codes, we call the two types of nodes \emph{variable nodes} and \emph{check nodes}, respectively.

SC decoding can be greatly simplified for the BEC as follows. All messages belong to an alphabet of cardinality three, which we define to be $\{-\infty,0,+\infty\}$. The symbol $0$ denotes an erasure. For a check node, the update rule consists of taking the product of the signs of the incoming messages. For a variable node, if we define $+\infty - \infty = -\infty + \infty = 0$, the update rule becomes a simple addition.

\section{Faulty SC Decoding of Polar Codes}\label{sec:faulty}

\subsection{Tree Channel and Density Evolution}
In order to analyze the erasure probability for each $W^{(i)}_N(y_1^N,\hat{u}_1^{i-1}|u_i)$, we use the notion of a \emph{tree channel}~\cite{Hassani2012}. In order to calculate the LLR $L(y_1^N,\hat{u}_1^{i-1}|u_i)$ required to decode each bit $u_i$, the $N$ channel LLRs, are combined through a tree-like structure of height $n$ with $n+1$ levels of nodes. $L(y_1^N,\hat{u}_1^{i-1}|u_i)$ is found at level $n$ (which is the root of the tree), while the channel LLRs are found at level $0$. Let $b$ denote the right-MSB $(n+1)$-bit binary expansion of $i$. All nodes at level $j$ are variable nodes if $b_j = 1$, and check nodes if $b_j = 0$. 

Due to channel and decoder symmetry, we can assume that the all-zero codeword was transmitted~\cite{Hassani2012}, meaning that, since $\hat{u}_s$ is always equal to $0$, the variable node update rule becomes $m = m_1 + m_2$. Moreover, since the output of a BEC is never erroneous, messages with value $-\infty$ can not appear during the decoding process. Thus, under the all-zero codeword assumption for the BEC, at a variable node the outgoing message is an erasure when both of the incoming messages are erasures, while at a check node the outgoing message is an erasure when any of the two incoming messages is an erasure. When the variable node update rule is applied to two independent messages with erasure probability $\epsilon$, the erasure probability of the outgoing message is given by
\begin{align}
	T^{+}(\epsilon) & \triangleq \epsilon ^2.
\end{align}
Similarly, for the application of the check node we have
\begin{align}
	T^{-}(\epsilon) & \triangleq 1 - (1-\epsilon )^2 = 2\epsilon - \epsilon ^2.
\end{align}
Following~\cite{Arikan2009,Hassani2012}, we define the random process $\epsilon _j$ as
\begin{align}
	\epsilon _{j+1} = \left\{ \begin{matrix} T^{+}(\epsilon_j) & \text{w.p. }~1/2, \\ T^{-}(\epsilon_j) & \text{w.p. }~1/2,  \end{matrix} \right. \label{eq:nonfaulty}
\end{align}
where $\epsilon _0 = p$ is the erasure probability of the channel messages, which are found at level $0$ of each tree. It was shown in \cite{Arikan2009} that $\epsilon _n$ converges almost surely to a random variable $\epsilon _{\infty} \in \{0,1\}$, with $P(\epsilon _{\infty} = 0) = 1-\epsilon$.

\subsection{Fault Model and Density Evolution for Faulty Decoding}
We model faulty decoding as additional erasures in the decoder, which may be caused either by faulty message processing or by faulty message storage. These additional erasures can only happen on messages that are not already erased, and they happen independently of whether the message value is $+\infty$ or $-\infty$ and with probability $\delta \in (0,1)$.\footnote{It is easy to check that for $\delta = 0$ we get a non-faulty decoder, while for $\delta = 1$ all messages are always erasures leading to a completely faulty decoder. Thus, it is mainly interesting to study the decoder for $\delta \in (0,1)$.} Thus, at a variable node the total erasure probability is
\begin{align}
	T^{+}_{\delta}(\epsilon) & \triangleq \epsilon ^2 + (1-\epsilon ^2)\delta, \label{eq:Tvarfaulty}
\end{align}
while at a check node we have
\begin{align}
	T^{-}_{\delta}(\epsilon) & \triangleq 2\epsilon - \epsilon ^2 + (1-2\epsilon + \epsilon ^2)\delta. \label{eq:Tcheckfaulty}
\end{align}
We can re-define the random process $\epsilon _j$ using \eqref{eq:Tvarfaulty} and \eqref{eq:Tcheckfaulty} as
\begin{align}
	\epsilon _{j+1} = \left\{ \begin{matrix} T^{+}_{\delta}(\epsilon_j) & \text{w.p. }~1/2, \\ T^{-}_{\delta}(\epsilon_j) & \text{w.p. }~1/2,  \end{matrix} \right. \label{eq:faultyfull}
\end{align}
where again $\epsilon _0 = p$. 

\subsection{Polarization Does Not Happen}
We first show some properties of $T^{+}_{\delta}(\epsilon)$ and $T^{-}_{\delta}(\epsilon)$, which will be useful to prove the main result of this section and to interpret some of the numerical results of Section~\ref{sec:numerical}.
\begin{proposition}\label{prop:properties}
For $T^{+}_{\delta}(\epsilon)$ and $T^{-}_{\delta}(\epsilon)$, we have
\begin{enumerate}
	\setlength{\itemsep}{3pt}
	\item[(i)] $T^{+}_{\delta}(\epsilon) < \epsilon ,~\forall \epsilon \in \left(\frac{\delta}{1-\delta},1\right)$,
	\item[(ii)] $T^{+}_{\delta}(\epsilon) > \epsilon ,~\forall \epsilon \in \left[0,\frac{\delta}{1-\delta}\right)$
	\item[(iii)] $T^{-}_{\delta}(\epsilon) > \epsilon ,~\forall \epsilon \in \left[0,1\right)$.
\end{enumerate}
\end{proposition}
\begin{IEEEproof}
For $T^{+}_{\delta}(\epsilon)$, we have
\begin{align}
	\epsilon ^2 + (1-\epsilon ^2)\delta & < \epsilon \Leftrightarrow \\
	(1-\delta)\epsilon ^2 - \epsilon + \delta & < 0.
\end{align}
The roots of $(1-\delta)\epsilon^2 - \epsilon + \delta = 0$ are $\epsilon = 1$ and $\epsilon = \frac{\delta}{1-\delta}$. Since $(1-\delta) > 0$, which is the coefficient of $\epsilon^2$, the sign of the function between the two roots will be negative and (i) follows. Moreover, using the same argument, we have $T^{+}_{\delta}(\epsilon) > \epsilon,~\forall \epsilon \in \left[0,\frac{\delta}{1-\delta}\right)$, so (ii) follows. For (iii) we have
\begin{align}
	2\epsilon - \epsilon ^2 + (1-2\epsilon + \epsilon^2)\delta & > \epsilon \Leftrightarrow \\
	(1-\epsilon)\left( \epsilon + (1-\epsilon)\delta\right) & > 0.
\end{align}
which indeed holds for any $\epsilon \in [0,1)$.
\end{IEEEproof}
\begin{proposition}\label{prop:fixed}
The fixed points of $T^{+}_{\delta}(\epsilon)$ are $\epsilon = 1$ and $\epsilon = \frac{\delta}{1-\delta}$. The unique fixed point of $T^{-}_{\delta}(\epsilon)$ for $\epsilon \in [0,1]$ is $\epsilon = 1$.
\end{proposition}
\begin{IEEEproof}
The above proposition can easily be shown by solving $T^{+}_{\delta}(\epsilon) = \epsilon$ and $T^{-}_{\delta}(\epsilon) = \epsilon$ for $\epsilon$, respectively, and noting that one solution of $T^{-}_{\delta}(\epsilon) = \epsilon$ is negative.
\end{IEEEproof}
Moreover, the following result about the process $\epsilon _j$ gives us some first insight into the effect that the faulty decoder has on the decoding process.
\begin{proposition}
The process $\epsilon _j,~j = 0,1,\hdots,$ defined in \eqref{eq:faultyfull} is a submartingale.
\end{proposition}
\begin{IEEEproof}
Since $\epsilon _j$ is bounded, it holds that $\mathbb{E}(|\epsilon _j|) < \infty$. Moreover we have
\begin{align}
	\mathbb{E}(\epsilon _{j+1}|\epsilon _j) & = \frac{1}{2}\left(T^{+}_{\delta}(\epsilon_j) + T^{-}_{\delta}(\epsilon_j)\right) \\
											& = \frac{1}{2}\left((1-\epsilon_j ^2)\delta + 2\epsilon_j + (1-2\epsilon_j + \epsilon_j ^2)\delta)\right) \\
											& = \epsilon_j + (1-\epsilon_j)\delta \geq \epsilon _j.
\end{align}
\end{IEEEproof}
Specifically, this tells us that, contrary to \cite{Arikan2009}, the overall erasure probability\footnote{Equivalently for the BEC, the mutual information and the Bhattacharyya parameter.} is not preserved by $T^{+}_{\delta}(\epsilon)$ and $T^{-}_{\delta}(\epsilon)$. So, even if fully reliable transmission were possible in the limit of infinite blocklength, this would come at the cost of a rate loss, so the polar code would not be capacity achieving. Unfortunately, as the following theorem asserts, fully reliable transmission under faulty decoding is not possible.

\begin{theorem}\label{thm:nopol}
Let $\mathcal{S}$ denote the sample space of the process $\epsilon_j$ and let $\epsilon_j(s),~s \in \mathcal{S},$ denote a specific realization of $\epsilon_j$. Polarization does not happen under faulty SC decoding for the BEC in the sense that $\nexists s \in \mathcal{S}$ such that $\epsilon_j(s) \stackrel{j \rightarrow \infty}{\longrightarrow} 0$. 
\end{theorem}
\begin{IEEEproof}
To see this, it is sufficient to observe that $T^{+}_{\delta}(\epsilon) \geq \delta$ and $T^{-}_{\delta}(\epsilon) \geq \delta$ for any $\epsilon \in [0,1]$. A more detailed proof is provided in the Appendix.
\end{IEEEproof}

\section{Numerical results}\label{sec:numerical}
So far, we have shown that fully reliable communication under faulty SC decoding for the BEC is unfortunately impossible. However, in practice fully reliable communication is typically not required. Thus, it is interesting to study the behavior of the SC decoder under faulty decoding and explore what \emph{is} in fact possible. To this end, in this section we provide some numerical results to explore the process $\epsilon _j$, as well as the FER performance of polar codes constructed based on this process.

\begin{figure}
	\centering
	\includegraphics[width=\figwidth\textwidth]{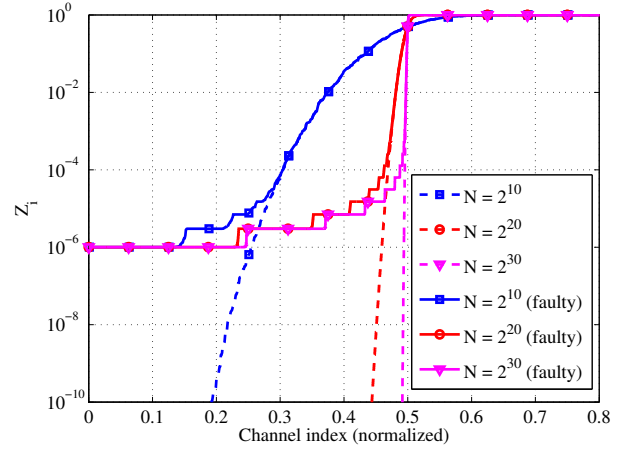}
	\caption{Sorted $Z_i$ values for polar codes of length $N = 2^{10}, 2^{20}, 2^{30},$ designed for the BEC$(0.5)$ under faulty SC decoding with $\delta = 10^{-6}$.}\label{fig:Z}
\end{figure}

\subsection{``Polarization'' Process $\epsilon _j$}\label{sec:numericalpolarization}
The Bhattacharyya parameters $Z_i,~i=1,\hdots,N,$ of a polar code of length $N$ for transmission over a BEC$(p)$ correspond to all possible realizations of $\epsilon _n$ for $\epsilon _0 = p$ and for a given $\delta$, where $n = \log N$. In Fig.~\ref{fig:Z}, we present the (sorted) values of $Z_i,~i = 1,\hdots,N,$ for polar codes of length $N = 2^{10},2^{20},2^{30},$ designed for the BEC$(0.5)$ under faulty SC decoding with $\delta = 10^{-6}$. We observe that we always have $Z_i \geq \frac{\delta}{1-\delta} \triangleq \epsilon ^*$, which is not surprising since from Proposition~\ref{prop:fixed} we know that that $\epsilon ^*$ is a fixed point of $T^{+}_{\delta}(\epsilon)$. However, $\epsilon ^*$ is not a fixed point of $T^{-}_{\delta}(\epsilon)$ (whereas $1$ is a fixed point for both), resulting in the staircase-like structure of Fig.~\ref{fig:Z}. 

The process $\epsilon _j$ is a bounded submartingale, so it converges almost surely to some limiting random variable $\epsilon _{\infty}$. Indeed, in our numerical studies we observe that, as $N$ is increased, the staircase structure becomes more pronounced and seems to converge to a limit. Unfortunately, we have not been able to identify that limit.

\subsection{Frame Erasure Rate}

\begin{figure}
	\centering
	\includegraphics[width=\figwidth\textwidth]{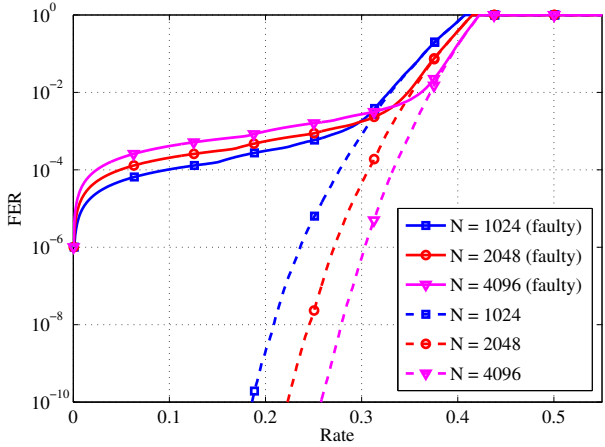}
	\caption{FER for polar codes of lengths $N = 1024,2048,4096,$ designed for the BEC$(0.5)$ with $\delta = 10^{-6}$.}\label{fig:FERvsR}
\end{figure}

Let $P_f(N,R)$ denote the frame erasure rate (FER) of a rate-$R$ polar code of length $N$. 
It was shown in \cite{Bastani2013} that for the BEC (under reliable decoding) we have
\begin{align}
	P_f(N,R) \approx \sum_{i \in \mathcal{A}} Z_i. \label{eqn:FER}
\end{align}
It is not clear whether the proof of \cite{Bastani2013} can be immediately extended to the faulty decoding case, but we nevertheless use \eqref{eqn:FER} as a proxy for the FER. In Fig.~\ref{fig:FERvsR}, we present the evaluation of $P_f(N,R)$ as a function of $R$ and for $N = 1024,2048,4096,$ for a faulty SC decoder with $\delta = 10^{-6}$ and transmission over the BEC$(0.5)$. We also present the FER under non-faulty decoding for comparison. Strikingly, over a wide range of rates, the FER under SC decoding actually increases when the blocklength is increased. This can be explained if we recall that $Z_i \geq \delta$. Thus, by increasing the blocklength while keeping the rate fixed, we are increasing the number of terms in \eqref{eqn:FER}, and since some of these terms do not decrease beyond some point, the value of the sum can increase.

\section{Unequal Error Protection}\label{sec:protection}

The SC decoder can be implemented as a tree of processing elements (PEs), which apply the update rules and also store the resulting messages~\cite{Leroux2011}. In essence, this tree of PEs is the implementation of a tree channel of depth $n$ in hardware, containing $n+1$ levels of PEs. At level $j$, we need $2^{n-j}$ PEs, so the total number of PEs required by a decoder is
\begin{align}
	N_{\text{PE}}	& = \sum _{j=0}^n2^{n-j} = 2^{n+1} -1 = 2N-1.
\end{align}

As mentioned in Section~\ref{sec:introduction}, standard methods employed to enhance the fault tolerance of circuits, such as using larger transistors or circuit-level error correcting codes, are costly in terms of both area and power if the whole circuit needs to be protected. However, not all levels in the tree of PEs are of equal importance, meaning that it may suffice to employ \emph{partial protection} of the decoder against hardware-induced errors. In fact, as Theorem~\ref{thm:uneqprot} asserts, a careful application of such a protection method allows polarization to happen even in a faulty decoder while protecting only a constant fraction of the total decoder PEs.

\begin{figure}
	\centering
	\includegraphics[width=\figwidth\textwidth]{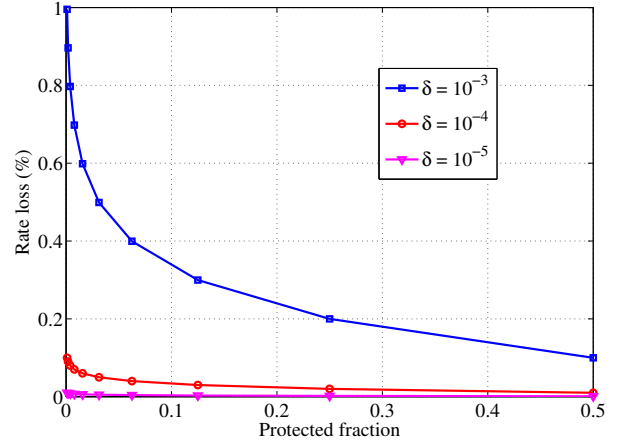}
	\caption{Rate loss $\Delta R (\delta,\epsilon,n_{\text{u}})$ (as a percentage of the capacity $C = 1-\epsilon$) for $n_{\text{u}} = 1,\hdots,10,$ and $\delta = 10^{-3},10^{-4},10^{-5}$. The channel is a BEC$(0.5)$.}\label{fig:RateLoss}
\end{figure}

Let $n_\text{p}$ denote the number of levels that are protected, starting from level $n$ of the tree (i.e., the root) and going towards the leaves. We assume that for these $n_\text{p}$ levels we have $\delta = 0$. Let $N_{\text{p}}$ denote the total number of protected PEs, where 
\begin{align}
	N_{\text{p}} = \left\{ \begin{matrix} \sum _{j=0}^{n_{\text{p}}-1}2^{j} = 2^{n_{\text{p}}}-1, & n_{\text{p}} > 0,\\ 0, & n_{\text{p}} = 0. \end{matrix} \right.
\end{align}
If we set $n_{\text{p}} = (n+1) - n_{\text{u}}$, where $n_{\text{u}} > 0$ is a \emph{fixed} number of unprotected levels, then the fraction of the decoder that is protected converges to a constant as $n$ grows. Indeed, we have
\begin{align}
	\lim _{n \rightarrow \infty}\frac{N_{\text{p}}}{N_{\text{PE}}} & = \lim _{n \rightarrow \infty}\frac{2^{(n + 1) - n_{\text{u}}}-1}{2^{n+1}-1} = 2^{-n_{\text{u}}}.
\end{align}
In this case, the process $\epsilon_j$ can be rewritten as
\begin{align}
	\epsilon _{j+1} = \left\{ \begin{matrix} 
															T_{\delta}^{+}(\epsilon_j), & \text{w.p. } 1/2, & \multirow{2}{*}{if $j = 0,\hdots,n_{\text{u}}-1$,} \\ 
															T_{\delta}^{-}(\epsilon_j), & \text{w.p. } 1/2, &  \\
															T^{+}(\epsilon_j), & \text{w.p. } 1/2, & \multirow{2}{*}{if $j = n_{\text{u}},\hdots,n$.} \\
															T^{-}(\epsilon_j), & \text{w.p. } 1/2, & 
														\end{matrix} \right. \label{eq:protej}
\end{align}
The following theorem asserts that the protection of a constant fraction of the decoder is sufficient to ensure that polarization happens as $n$ grows.
\begin{theorem}\label{thm:uneqprot}
Setting $n_{\text{p}} = (n+1) - n_{\text{u}}$ for any fixed $n_{\text{u}}$ suffices to ensure that $\epsilon _j$ converges almost surely to a random variable $\epsilon _{\infty} \in \{0,1\}$. However, the unprotected levels result in a rate loss $\Delta R (\delta,\epsilon,n_{\text{u}})$, in the sense that $P(\epsilon _{\infty} = 0) = 1 - \epsilon - \Delta R (\delta,\epsilon,n_{\text{u}})$.
\end{theorem}
\begin{IEEEproof}
The process $\epsilon_{j}$ as defined in \eqref{eq:protej} is a submartingale for $j < n_{\text{u}}$, but it becomes a martingale for $j \geq n_{\text{u}}$. Thus, for $j \geq n_{\text{u}}$ we have $\mathbb{E}(\epsilon _{j}) = \mathbb{E}(\epsilon _{n_{\text{u}}})$. Using the arguments from \cite{Arikan2009}, we can show that $\epsilon _j$ converges almost surely to a random variable $\epsilon _{\infty} \in \{0,1\}$ with $P(\epsilon _{\infty} = 0) = 1-\mathbb{E}(\epsilon _{n_{\text{u}}}) \leq 1 - \epsilon$. Equivalently, $P(\epsilon _{\infty} = 0) = 1 - \epsilon - \Delta R (\delta,\epsilon,n_{\text{u}})$ for $\Delta R (\delta,\epsilon,n_{\text{u}}) = \mathbb{E}(\epsilon_{n_{\text{u}}}) - \epsilon$, where $\mathbb{E}(\epsilon_{n_{\text{u}}})$ can be evaluated numerically via density evolution. 	
\end{IEEEproof}

\begin{figure}
	\centering
	\includegraphics[width=\figwidth\textwidth]{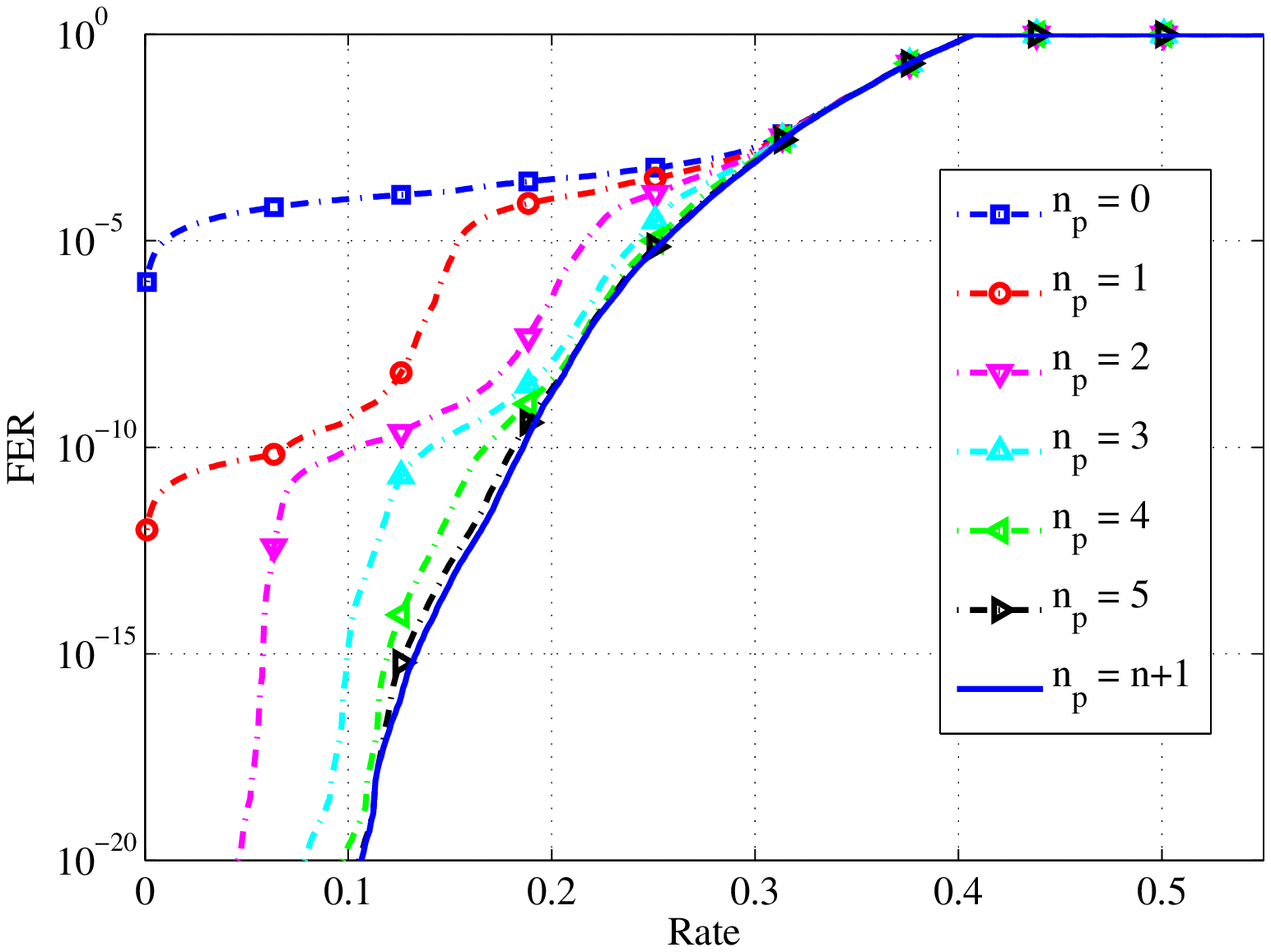}
	\caption{FER for a polar code of length $N = 1024$ designed for the BEC$(0.5)$ under faulty SC decoding with $\delta = 10^{-6}$ and $n_{\text{p}} = 0, \hdots, 5,$ protected decoding levels.}\label{fig:FERvsStageProtection}
\end{figure}

\begin{figure}
	\centering
	\includegraphics[width=\figwidth\textwidth]{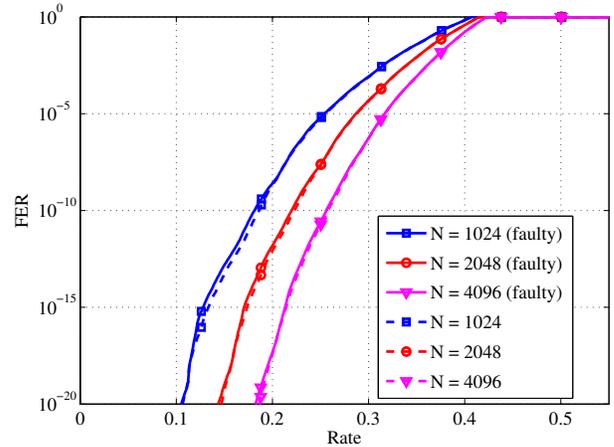}
	\caption{FER for polar codes of length $N = 1024, 2048, 4096,$ designed for the BEC$(0.5)$ under faulty SC decoding with $\delta = 10^{-6}$ and $n_{\text{p}} = n-5$ protected decoding levels.}\label{fig:FERvsStageProtectionN}
\end{figure}

Theorem \ref{thm:uneqprot} implies that, when partial protection of the decoder is employed, polar codes are still not capacity achieving, but they can nevertheless be used for reliable transmission at any rate $R$ such that $R < 1-\epsilon-\Delta R (\delta,\epsilon,n_{\text{u}})$. The rate loss $\Delta R (\delta,\epsilon,n_{\text{u}})$ is presented in Fig.~\ref{fig:RateLoss} as a function of $n_{\text{u}}$ for some values of $\delta$ for a code designed for the BEC$(0.5)$. Moreover, the effect of the partial protection for a finite length code is illustrated in Fig.~\ref{fig:FERvsStageProtection}, where we present $P_f(N,R)$ for $N = 1024$ and $\delta = 10^{-6}$ when $n_{\text{p}} = 0,\hdots,5,$ levels of the tree are protected. We observe that protecting only the root node already improves the performance significantly, especially for the lower rates. When $n_{\text{p}} = 5$, the performance of the faulty SC decoder is almost identical to the non-faulty decoder and it is remarkable that this performance improvement is achieved by protecting only $\frac{N_{\text{p}}}{N_{\text{PE}}} = \frac{31}{2047} \approx 1.5\%$ of the decoder. Moreover, in Fig.~\ref{fig:FERvsStageProtectionN}, we present $P_f(N,R)$ for $N = 1024, 2048, 4096$ and $\delta = 10^{-6}$ with $n_{\text{p}} = n - 5$, so that the protected part for each $N$ is fixed to approximately $1.5\%$ of the decoder. We observe that, contrary to the results of Section~\ref{sec:numerical}, increasing the blocklength actually decreases $P_f(N,R)$, as in the case of the non-faulty decoder.

\section{Conclusion}\label{sec:conclusion}
In this paper, we studied faulty SC decoding of polar codes for the BEC, where the hardware-induced errors are modeled as additional erasures within the decoder. We showed that, under this model, fully reliable communication is not possible. Moreover, we presented numerical frame erasure rate results to explore what \emph{is} possible under faulty SC decoding. Finally, we proposed an error protection scheme which can significantly improve the performance of a faulty SC decoder by protecting as little as $1.5\%$ of the decoder.

\section*{Acknowledgment}
The authors would like to thank the anonymous reviewers for their helpful comments. This work was kindly supported by the Swiss NSF under Project ID 200021\_149447.

\appendix

\begin{IEEEproof}[Proof of Theorem~\ref{thm:nopol}]
Any $\epsilon_j(s),~s \in \mathcal{S},~j > 0,$ results from repeated applications of $T^{+}_{\delta}(\epsilon)$ and $T^{-}_{\delta}(\epsilon)$ to $\epsilon _0(s) = p$. From this point on, we denote $\epsilon_j(s)$ by $\epsilon_j$ for simplicity. From Proposition~\ref{prop:properties}, we know that $T^{-}_{\delta}(\epsilon_j)$ is strictly increasing for $\epsilon _j \in [0,1)$, and that $T^{+}_{\delta}(\epsilon_j)$ is strictly decreasing for $\epsilon _j \in \left(\frac{\delta}{1-\delta},1\right)$ and strictly increasing for $\epsilon _j \in \left[0,\frac{\delta}{1-\delta}\right)$. So, in order to show that $\epsilon _j$ can not become arbitrarily small, it suffices to show that $T^{+}_{\delta}(\epsilon_j)$ can not decrease the value of $\epsilon _j$ beyond some strictly positive value when $\epsilon _j \in \left(\frac{\delta}{1-\delta},1\right)$. Indeed, $T^{+}_{\delta}(\epsilon_j) \geq \delta$ for any $\epsilon _j \in [0,1]$, so the claim holds. 
\end{IEEEproof}


\ifCLASSOPTIONcaptionsoff
  \newpage
\fi

\end{document}